\documentstyle[titlepage,12pt]{article}
\begin{document}
\title{Comment on thickness-corrections to Nambu-wall}
\vspace{5in}
\author{A.L.Larsen\thanks{E-mail: allarsen@nbivax.nbi.dk}\\
Nordita, Blegdamsvej 17, DK-2100 Copenhagen \O, Denmark}
\maketitle
\begin{abstract}
We comment on some calculations concerning the finite-thickness
corrections to the (generalized) Nambu action
for a curved domain wall in Minkowski space. Contrary to a recent claim in
the literature, we find no first order corrections in the wall-thickness,
and only one second order correction proportional to the Ricci curvature of
the wall. These results are obtained by consistently expanding the action
and the equations of motion for the scalar field.
\end{abstract}
\newpage
Cosmic strings and domain walls play an important role in cosmological models
concerning phase transitions in the early universe, galaxy formation,
etc...[1,2]. Although the domain wall is a topological defect of one dimension
higher than the string, it is in some senses easier to describe: From the
field theoretic point of view a domain wall can be formed in theories of just
one real scalar field whereas the strings involve more complicated
field-configurations, and from a (generalized) world-sheet point of view the
domain wall, being a hypersurface, has no torsion and only one component of
the mean curvature and the Gaussian curvature.

The dynamics of walls is usually described by a generalized Nambu action
corresponding to the assumption that the (other) dimensions of the wall are
much greater than the thickness. During the last few years, however, there
has been some interest in calculating the finite thickness corrections to the
Nambu action. The main results are due to Gregory, Haws and Garfinkle [4,5,6],
who analyzed the coupled Einstein-scalar equations for a thick gravitating
wall, using methods from differential geometry. Much more recently the same
problem (although in flat spacetime) was attacked by Silveira and Maia [7]
using a somewhat simpler approach. After making a suitable {\it Ansatz} for
the scalar field in the vicinity of the wall, they expanded the action and
the equations of motion around a so-called "locally non-plane solution"
in powers of the wall-thickness $\epsilon$. We find however, that their
calculations are inconsistent: First of all the results of Ref. 7 depend
crucially on whether the thickness expansions are made directly in
the action or in the
equations of motion (and this is not because of the {\it Ansatz} for the
scalar field!). Secondly the final result of Ref. 7 for the effective
action of the wall is plagued with the disease that partial integrations
completely change the power counting and therefore mixe the different terms
of different powers in the expansion.

The purpose of this note is to present a consistent derivation of the
effective action for the curved thick wall, using the simple
approach of Ref. 7.

The starting point is the action for a real scalar field in Minkowski space:
\begin{equation}
{\cal S}=\int[-\frac{1}{2}\eta^{\mu\nu}\partial_\mu\phi\partial_\nu\phi-
V(\phi)]d^4z,
\end{equation}
where $\eta_{\mu\nu}=diag(-1,1,1,1)$ and:
\begin{equation}
V(\phi)=\lambda(\phi^2-v^2)^2.
\end{equation}
It is well-known that this model enjoys wall-solutions [2,3], and the idea
is now to introduce a coordinate system based on the wall-surface. We thus
write:
\begin{equation}
z^\mu(\sigma^A,\xi)=x^\mu(\sigma^A)+\xi n^\mu(\sigma^A),
\end{equation}
where $x^\mu(\sigma^A)$ describes the wall-surface embedded in Minkowski space,
$\sigma^A$ $(A=0,1,2)$ are the 3 intrinsic coordinates on the wall-surface and
$n^\mu(\sigma^A)$ is the vector normal to the surface:
\begin{equation}
\eta_{\mu\nu}n^\mu n^\nu=1,\hspace*{5mm}\eta_{\mu\nu}n^\mu x^\nu_{,A}=0.
\end{equation}
The metric in the new coordinate system is:
\begin{equation}
ds^2=\eta_{\mu\nu}dz^\mu dz^\nu=g_{AB}d\sigma^A d\sigma^B+
2g_{A\xi}d\sigma^A d\xi+g_{\xi\xi}d\xi^2,
\end{equation}
where:
\begin{eqnarray}
&g_{AB}=\eta_{\mu\nu}x^\mu_{,A}x^\nu_{,B}+2\xi\eta_{\mu\nu}n^\mu_{,A}x^\nu_{,B}
+\xi^2\eta_{\mu\nu}n^\mu_{,A}n^\nu_{,B},&\nonumber\\
&g_{\xi\xi}=1,\hspace*{5mm}g_{A\xi}=g_{\xi A}=0.&
\end{eqnarray}
The action (1) now reads:
\begin{equation}
{\cal S}=\int\sqrt{-g}[-\frac{1}{2}g^{\mu\nu}\partial_\mu\phi\partial_\nu\phi
-V(\phi)]d^3\sigma d\xi,
\end{equation}
where $g_{\mu\nu}$ is given by (6). The corresponding equations of motion are:
\begin{equation}
\frac{1}{\sqrt{-g}}\partial_\mu(\sqrt{-g}g^{\mu\nu}\partial_\nu\phi)-
\frac{\partial V}{\partial\phi}=0.
\end{equation}
We will consider a domain wall solution $\phi$ and expand the equations of
motion (8) in powers of the thickness $\epsilon$. We will assume that close to
the wall the field $\phi$ depends only on the transverse coordinate $\xi$. The
field itself therefore is expanded as:
\begin{equation}
\phi=\phi_{(0)}(\xi)+\phi_{(1)}(\xi)+\phi_{(2)}(\xi)+...,
\end{equation}
with $\phi_{(1)}$ of order $\epsilon$, $\phi_{(2)}$ of order $\epsilon^2$, etc.
Furthermore we will assume that the derivative in the normal direction is
one order in $\epsilon$ lower than the derivatives in the tangential
directions. This latter assumption was not included in the approach of
Silveira and Maia [7], but it is in fact a necessary assumption,
not only for the
consistency of the formal calculations; if the expansion in the thickness
is to be meaningful at all, the wall-thickness is supposed to be
small (compared to the curvature radius of the wall), which
implies fast fall-off properties of $\phi$ [4,5,6]. The various objects are
then assigned the following orders in $\epsilon$:
\begin{eqnarray}
&\phi_{(0)}\sim 1,\hspace*{5mm}\phi_{(1)}\sim\epsilon,\hspace*{5mm}
\phi_{(2)}\sim\epsilon^2,&\nonumber\\
&\xi\sim\epsilon,\hspace*{5mm}"\partial_A\sim 1",\hspace*{5mm}"\partial_\xi\sim
\epsilon^{-1}",&\nonumber\\
&\partial_\xi\phi_{(0)}\partial_\xi\phi_{(0)}\sim\epsilon^{-2},\hspace*{5mm}
V(\phi_{(0)})\sim\epsilon^{-2}.&
\end{eqnarray}
Using (6) we can expand $\sqrt{-g}$ in powers of $\xi$, which is then
automatically an expansion in $\epsilon$ [7]:
\begin{equation}
\sqrt{-g}=\sqrt{-G}[1+K_\xi\xi+\frac{1}{2}(K_\xi K_\xi-K_{\xi\xi})\xi^2+...].
\end{equation}
Here $G_{AB}$ is the induced metric on the wall-surface:
\begin{equation}
G_{AB}=\eta_{\mu\nu}x^\mu_{,A}x^\nu_{,B},
\end{equation}
$K_\xi$ is the mean curvature [8]:
\begin{equation}
K_\xi=G^{AB}\eta_{\mu\nu}n^\mu_{,A}x^\nu_{,B}=n^{\mu,A}x_{\mu,A},
\end{equation}
and $K_{\xi\xi}$ is the Gaussian curvature [8]:
\begin{equation}
K_{\xi\xi}=G^{AC}G^{BD}\eta_{\mu\nu}n^\mu_{,A}x^\nu_{,B}\eta_{\rho\sigma}
n^\rho_{,C}x^\sigma_{,D}=n^{\mu,A}n_{\mu,A}.
\end{equation}
At the 2 lowest orders in $\epsilon$ the expansion of the equations of motion
(8) leads to:
\begin{equation}
\epsilon^{-2}:\hspace*{1cm}\partial^2_\xi\phi_{(0)}-\frac{\partial V}{
\partial\phi}\mid_{(0)}=0,
\end{equation}
\begin{equation}
\epsilon^{-1}:\hspace*{1cm}\partial^2_\xi\phi_{(1)}+K_\xi\partial_\xi\phi_{(0)}
-\frac{\partial^2 V}{\partial\phi^2}\mid_{(0)}\phi_{(1)}=0.
\end{equation}
If we make instead the expansion directly in the action (7) we find at lowest
order:
\begin{equation}
{\cal S}_0=\int\sqrt{-G}[-\frac{1}{2}\partial_\xi\phi_{(0)}\partial_\xi
\phi_{(0)}-V(\phi_{(0)})]d^3\sigma d\xi,
\end{equation}
which is consistent with (15). At the next order:
\begin{equation}
{\cal S}_1=\int\sqrt{-G}[-\partial_\xi\phi_{(1)}\partial_\xi\phi_{(0)}-
\frac{\partial V}{\partial\phi}\mid_{(0)}\phi_{(1)}+K_\xi
(-\frac{1}{2}\partial_\xi\phi_{(0)}\partial_\xi\phi_{(0)}-V(\phi_{(0)}))\xi]
d^3\sigma d\xi.
\end{equation}
The first 2 terms in the square bracket vanish because of the lowest order
equation of motion (15), and also the $K_\xi$-term vanishes: The general
solution of (15) is a Weierstrass elliptic function [9] which however
reduces to a hyperbolic function when the appropriate boundary conditions
are invoked:
\begin{equation}
\phi_{(0)}=v\tanh(\sqrt{2\lambda}v\xi).
\end{equation}
The $K_\xi$-term of (18) now vanishes after integration over $\xi$ because of
anti-symmetry. It is important to stress here that $\phi_{(0)}$ of (19) is not
the plane wall solution usually discussed in the literature [2,3];
equation (19)
only expresses the local behaviour of the wall.

Going one order higher in $\epsilon$ we find:
\begin{eqnarray}
{\cal S}_2=\int\sqrt{-G}[\hspace*{-2mm}&-&\hspace*{-2mm}\frac{1}{2}\partial_\xi
\phi_{(1)}\partial_\xi\phi_{(1)}-
\frac{1}{2}\frac{\partial^2 V}{\partial\phi^2}\mid_{(0)}
\phi^2_{(1)}+K_\xi\phi_{(1)}\partial_\xi\phi_{(0)}\nonumber\\
\hspace*{-2mm}&+&\hspace*{-2mm}(K_\xi K_\xi-K_{\xi\xi})(-\frac{1}{2}
\partial_\xi\phi_{(0)}\partial_\xi\phi_{(0)}-V(\phi_0))\xi^2]d^3\sigma d\xi,
\end{eqnarray}
where the lowest order equation of motion (15) has been used. Variation with
respect to $\phi_{(1)}$ leads to (16) so in contrary to Silveira and Maia [7]
our actions ${\cal S}_0$ and ${\cal S}_2$ are consistent with the equations
of motion. What we are interested in however, is an effective action describing
the dynamics of the wall. Therefore, collecting ${\cal S}_0$ and ${\cal S}_2$
and integrating over $\xi$:
\begin{equation}
{\cal S}=\int\sqrt{-G}[\mu_0+\tilde{\mu}_2K_\xi+\mu_2(K_\xi K_\xi-
K_{\xi\xi})+...]d^3\sigma,
\end{equation}
where:
\begin{equation}
\mu_0=\int[-\frac{1}{2}\partial_\xi\phi_{(0)}\partial_\xi\phi_{(0)}-
V(\phi_{(0)})]d\xi,
\end{equation}
\begin{equation}
\tilde{\mu}_2=\frac{1}{2}\int\phi_{(1)}\partial_\xi\phi_{(0)}d\xi,
\end{equation}
\begin{equation}
\mu_2=\int\xi^2[-\frac{1}{2}\partial_\xi\phi_{(0)}\partial_\xi\phi_{(0)}
-V(\phi_{(0)})]d\xi.
\end{equation}
Note that both $\tilde{\mu}_2$ and $\mu_2$ are 2 orders in $\epsilon$ higher
than $\mu_0$ (because of (10)).
This is however not the final result. At lowest order in $\epsilon$ we find
the equation of motion for the wall:
\begin{equation}
\frac{1}{\sqrt{-G}}\partial_A(\sqrt{-G}G^{AB}\partial_B x^\mu)=0,
\end{equation}
which implies $K_\xi=0$ at lowest order. In this case equation (16) is easily
solved for $\phi_{(1)}$, but the only solution with the appropriate
boundary conditions is $\phi_{(1)}=0$. This means that $\tilde{\mu}_2=0$
so that the effective action for the wall becomes:
\begin{equation}
{\cal S}=\int\sqrt{-G}[\mu_0-\mu_2 R+...]d^3\sigma,
\end{equation}
where, according to the Gauss-Codazzi equation [8],
$R=-K_\xi K_\xi+K_{\xi\xi}$
is the Ricci curvature of the wall. This result is in agreement with Gregory
[6], but in disagreement with Silveira and Maia [7].

In conclusion we have calculated the lowest order correction to the Nambu
action for a curved domain wall following the simple approach of Silveira and
Maia [7]. We found however that when the thickness expansions are made
consistently in the action and the equations of motion the extra terms
found in Ref. 7 are absent, and therefore the original result of Gregory
[6] is correct. We believe that this reflects that when the {\it necessary}
assumptions for the method to work are made (i.e. $\phi$ depends only on
the transverse
coordinate and $\phi$ has suitably fast fall-off properties),
then there is no such thing as a
"locally non-plane solution" as introduced in Ref. 7, and that is why
the result of Gregory is correct.

\newpage
\centerline{\bf References}
\begin{enumerate}
\item T.W.B. Kibble, Phys. Reports 67 (1980) 183
\item A. Vilenkin, Phys. Reports 121 (1985) 263
\item Y.B. Zeldovich, I.Y. Kobzarev and L.B. Okun, Sov. Phys. JETP 40
      (1975) 1
\item D. Garfinkle and R. Gregory, Phys. Rev. D41 (1990) 1889
\item R. Gregory, D. Haws and D. Garfinkle, Phys. Rev. D42 (1990) 343
\item R. Gregory, Phys. Rev. D43 (1991) 520
\item V. Silveira and M.D. Maia, Phys. Lett. A174 (1993) 280
\item L.P. Eisenhart, Riemannian Geometry (Princeton University Press,
      fifth printing, 1964), chapter IV
\item M. Abramowitz and I.A. Stegun, Handbook of Mathematical functions
      (Dover Publications Inc, New York, ninth printing), chapter 18
\end{enumerate}
\end{document}